# A note on the use of equidistant contours for presenting scientific data

by Hans van Haren

Royal Netherlands Institute for Sea Research (NIOZ), P.O. Box 59, 1790 AB Den Burg, the Netherlands.
e-mail: hans.van.haren@nioz.nl

The passionate plea for the use of 'scientific colour maps'[1] misses some aspects in the visual presentation of scientific data. While a linear colour map based on scientific human colour perception is useful for the presentation of some images, like the three examples given of the topography of the earth, an apple and a passport photograph, scientific data are not presented. In this note it will be shown that there is more in scientific oceanographic data as they are presented in forms varying from historic equidistant contours, via a linear black-(grey)-white 'b&w' map, a linear colour map and a nonlinear colour map. From an objective perspective, equidistant contouring is the best means for presenting scientific information in a relatively unbiased way. Nonlinear colour maps may add information to that by highlighting certain aspects also by varying the colour range if needed. Such information is not available from linear colour maps. Finally, the aesthetic aspect of visual data presentation is discussed.

Scientific data analysis and presentation requires extensive training, including the visual art of looking. Qualitative information is obtained by what an image superficially shows, quantitative information may only be obtained to a certain extent and generally requires further computational analysis. However, to attract the reader to a scientific data image, the presentation is important, as scientific looking is culturally dependent[2], and may even in some cases be 'catchy'. Such catchy presentation was more difficult in the era when publications included graphics in b&w only. However, from those days an informative means remains in the use of equidistant contouring, with the notion that the value of distance between contours may be varied by the author, by which the reader may be led.

Here, a subset is considered of recently, in 2019, obtained high-resolution temperature data showing a 100-m tall breaking wave in the deep-ocean to demonstrate the various ways of conveying a message (Fig. 1). The deep-ocean area is known for strong 'internal wave' breaking above a large underwater seamount[3]. The equidistant black contours leave a lot white in Fig. 1a, but they clearly show more smooth wavelike motions in the upper half of the image, and rugged and multiple closed contours reflecting turbulent overturning in the lower half including a sharp front around half-time. The sharpness of the front and the variation in vertical temperature gradients may be inferred by measuring the distance between contours. These data are used to quantitatively study the transfer of internal wave energy to turbulent mixing, which is vital for the transport of matter in the deep-ocean. Without the contours, the linear b&w map in Fig. 1b shows the wave but misses a good number of fine details, and only with great difficulty small-scale overturns are discernable. By adding colour, the linear batlow map[1] in Fig. 1c shows somewhat more fine details but the image remains rather flat. The use of a nonlinear colour map as in Fig. 1d adds a more three-dimensional '3D' non-flat aspect to the image, which is important for understanding and spotting turbulent overturning details. Turbulence is an essentially 3D process, and may be presented as such by showing all the large and small overturning scales, even when measurements are made in 1D or 2D (counting time as dimension). However, the best effort is made by combining the equidistant contours with a modified nonlinear map (Fig. 2). The modification shifts the attention to the rim of the breaking of the wave, where the claw-like smaller overturning occurs, still 10-m in height.

The painting of cloudy skies was an art in itself in the 17$^{th}$ century, some painters truly mastered the conveying of 3D turbulence to a 2D canvas. That attracts scientists alike, as a turbulent cloud immediately stands out, see also the contrast with the weaker turbulent upper part in Fig. 2. Whilst the use of equidistant contours is of importance for maintaining the objectiveness of a presented image, better than any linear b&w or colour map, in some presentations its combination with a nonlinear colour map like in Fig. 2 adds value by emphasizing certain aspects, in this case of the particular physics process. This notion is confirmed by a colleague who is colour blind, a matter which has been given substantial attention[1] and for good reason. About 8% of the, mainly male, human population suffers some kind of colour deficiency. That is a large number. But why deprive them to a certain extent and the other 92% from more informative and more beautiful imagery? Something that intrigues or is attractive keeps the attention longer and invites for a deeper inspection, in comparison with something that is flat and dull. More attention is the basis of the art of looking[4,5] and thus of scientific analysis.

As any painter knows, colour palettes are meant to emphasize certain aspects, so that the human brain differently interprets. This has also been shown from scientific human perception reasoning[1]. Arguably every palette does that, no matter how sequential, linear or nonlinear, thereby also retaining a subjective aspect to a certain extent. See the different aspects impressions in the different panels of Fig. 1. For least biased scientific data presentation, any colour map should be overlaid with equidistant contours. Using the combination of equidistant contouring and nonlinear colour mapping adds value to both data presentations and guides the reader what the authors want to present, in this case the truly 3D character of a breaking deep-ocean wave.

**Acknowledgements**

I thank Marek Stastna (Univ Waterloo, Ontario, Canada) for the darkjet colour that 'may fit your data well'. In memory of the late Jef Zimmerman who taught me to look at data.

**Statement**

The author declares no conflict of interest.

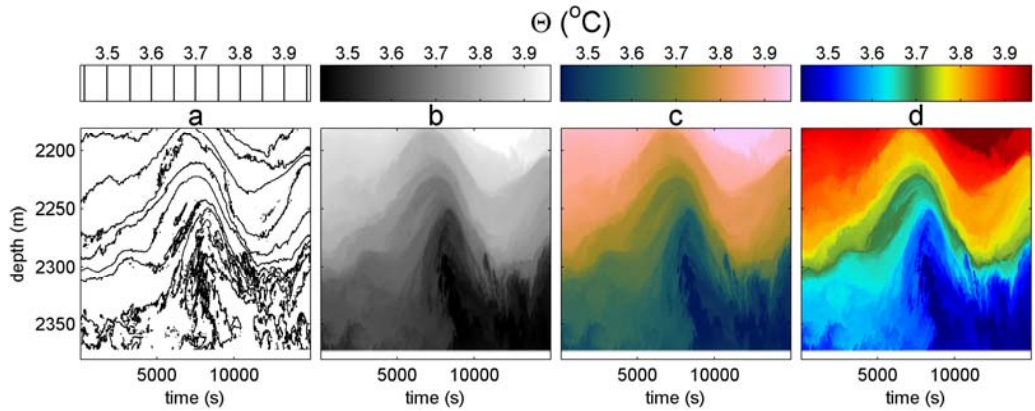

**Figure 1**. Four different presentations of high-resolution temperature observations from a 100 m tall breaking internal wave above a slope of a large underwater seamount. The 200 m vertical 15000 s time images represent about 2 million data points from 150 independent sensors sampling at a rate of 1 Hz. The data are a subset from a taut-wire deployment above Mount Josephine located 250 km West from South-Portugal in 2019. In all images, the (compressibility corrected) temperature range is [3.44 and 3.96]°C and the seafloor is at the depth-level of the horizontal axis. The four different displaying means are: (a) Equidistant contours every 0.05°C from 3.45 to 3.95°C. (b) B&w (via grey) linear map. (c) Batlow linear colour map[1]. (d) Modified rainbow 'darkjet' nonlinear colour map.

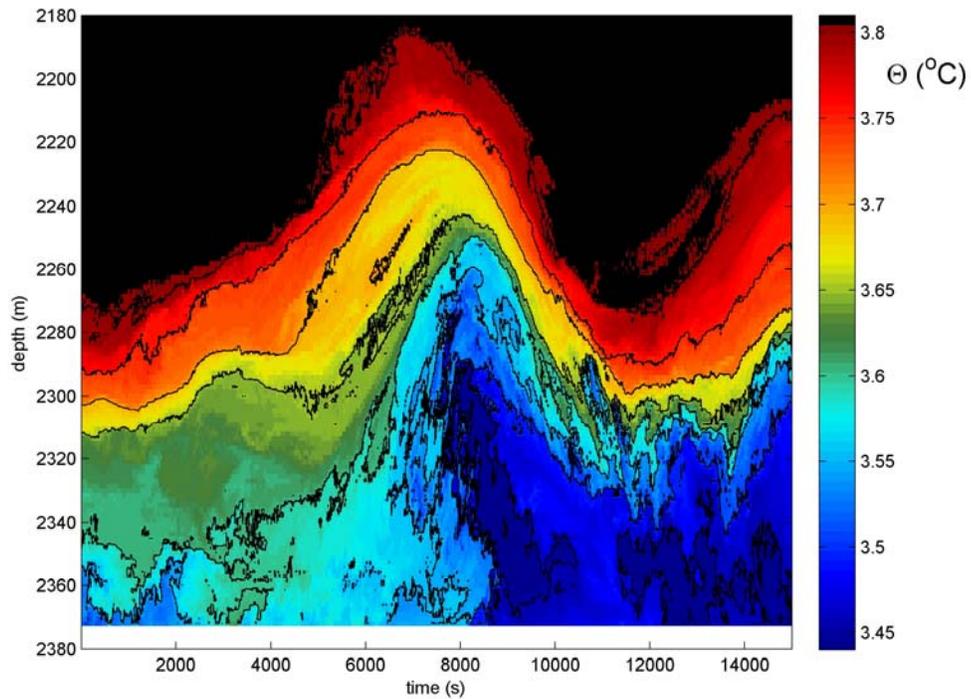

**Figure 2**. The same data window as in Fig. 1, emphasizing the turbulently breaking character of a big underwater surf, as represented using a further modified rainbow-darkjet colourmap overlain with equidistant contours every 0.05°C in black across a slightly modified temperature range. This image attracts the eye into the 3D-aspect of full turbulence. It focuses on the transition from smooth contours to the ruggedly deformed contours around the steep vertical front (around 8000 s) and the subsequent 'clawing' of minor turbulent overturning (around 2300 m and between 10000 and 14000 s). Also note the fluffy small turbulent bursting into the dark background on top of the wave crest.